 \documentclass[preprint]{aastex631}

\graphicspath{{./}{figures/}}

\begin{document}

\title{FAST search for circumstellar atomic hydrogen. II. Is BD+30$\arcdeg$3639 an interacting planetary nebula?}

\correspondingauthor{Yong Zhang}
\email{zhangyong5@mail.sysu.edu.cn}

\author[0000-0002-2762-6519]{Xu-Jia Ouyang}
\affiliation{School of Physics and Astronomy, Sun Yat-sen University, 2 Daxue Road, Tangjia, Zhuhai, Guangdong Province,  PR China}

\author[0000-0002-1086-7922]{Yong Zhang}
\affiliation{School of Physics and Astronomy, Sun Yat-sen University, 2 Daxue Road, Tangjia, Zhuhai, Guangdong Province,  PR China}
\affiliation{CSST Science Center for the Guangdong-Hongkong-Macau Greater Bay Area, Sun Yat-Sen University, Guangdong Province, PR  China}
\affiliation{Laboratory for Space Research, The University of Hong Kong, Hong Kong, PR  China}

 \author[0000-0002-3171-5469]{Albert Zijlstra}
 \affiliation{Department of Physics and Astronomy, The University of Manchester, Manchester M13 9PL, UK}
 \affiliation{Laboratory for Space Research, The University of Hong Kong, Hong Kong, PR China}

 \author[0000-0002-4428-3183]{Chuan-Peng Zhang}
 \affiliation{National Astronomical Observatories, Chinese Academy of Sciences, Beijing 100101, PR China}
 \affiliation{CAS Key Laboratory of FAST, National Astronomical Observatories, Chinese Academy of Sciences, Beijing 100101, PR China}

 \author[0000-0003-3324-9462]{Jun-ichi Nakashima}
 \affiliation{School of Physics and Astronomy, Sun Yat-sen University, 2 Daxue Road, Tangjia, Zhuhai, Guangdong Province,  PR China}
 \affiliation{CSST Science Center for the Guangdong-Hongkong-Macau Greater Bay Area, Sun Yat-Sen University, Guangdong Province, PR China}
                
 \author[0000-0002-2062-0173]{Quentin A Parker}
 \affiliation{Faculty of Science, The University of Hong Kong, Chong Yuet Ming Building Pokfulam Road, Hong Kong, PR China}
 \affiliation{Laboratory for Space Research, The University of Hong Kong, Hong Kong, PR China}
 \affiliation{CSST Science Center for the Guangdong-Hongkong-Macau Greater Bay Area, Sun Yat-Sen University, Guangdong Province, PR China}



\begin{abstract}

The young, compact, very high surface brightness but low excitation planetary nebula (PN) BD+30$^{\circ}$3639 is one of the
very few PNe that have been reported to exhibit the
21\,cm \ion{H}{1} emission line. As part of a long-term 
programme to search for circumstellar atomic hydrogen,
we observed the 21\,cm feature toward BD+30$^{\circ}$3639 
with  the Five-hundred-meter Aperture Spherical radio Telescope (FAST). 
Assuming a direct association between the PN and the detected \ion{H}{1} emission, these new 
observations show that this surrounding emission is significantly 
more spatially extended than indicated by previous 
interferometric observations, and can be resolved into two
velocity components. The estimated \ion{H}{1} mass is larger
than 100\,M$_\sun$, invalidating an origin from the host star itself or its ejecta for the emitting material. 
We discuss the possibility that the extended  \ion{H}{1} emission 
stems from the interstellar medium (ISM) swept out over time by the stellar wind.
Moreover, we report tentative detections of \ion{H}{1} absorption features lying near and blueward of 
the systemic velocity of
this PN, which are probably from a stalled asterosphere at the outer boundary of the expanding ionized region. 
The mass of the gas producing the \ion{H}{1} absorption  
is insufficient to solve the so-called `PN missing mass problem'. 
We demonstrate the capability of FAST to investigate the
interaction process between a PN and the surrounding ISM.

\end{abstract}

\keywords{Planetary nebulae (1249) --- Single-dish antennas (1460) --- Circumstellar envelopes (237) --- Stellar mass loss (1613) --- Interstellar atomic gas (833)}

\section{Introduction} \label{sec:intro}

As the descendant of low- and intermediate-mass stars (0.8 to 8\,${\rm M}_\sun$),
planetary nebulae (PNe) are important tools for understanding the history of stellar winds and the life cycle of materials in a Galaxy \citep{2022PASP..134b2001K}. However, extensive PN observations show that the total mass of the ionized material and the central star is only up to 1.5 M$_\sun$ \citep[e.g.,][]{2012A&A...541A.112K}, much lower than the theoretical upper limit. 
This is the so-called `PN missing mass problem' \citep{1994PASP..106..344K}. 
CO and H$_2$ observations indicate that PNe have molecular envelopes with a mass from
$10^{-2}$ to a few M$_\sun$ \citep{1996A&A...315..284H, DeMarco2022}, while
infrared observations suggest that PNe have low dust-to-gas mass ratios  \citep[$\sim0.01$,][]{2007MNRAS.381..117P,2020MNRAS.491..758A}, 
suggesting that the missing mass of PNe is not in either molecules or dust grains. Furthermore, only a modest number of PNe have detected so-called `AGB' haloes where at least some of this missing mass is located \citep[see, e.g.,][]{2003MNRAS.340..417C}, so the overall problem remains.
Another potentially substantial nebular component is atomic hydrogen (\ion{H}{1}),
especially in ionization-bound PNe \citep[see, e.g.,][for a review]{2022PASP..134b2001K}.

\ion{H}{1} has been detected through the 21\,cm emission in several circumstellar shells surrounding evolved stars \citep[$\geq$1 pc;][]{2006AJ....132.2566G}, demonstrating that
the \ion{H}{1} emission could serve as a tracer of the early history of PN mass loss. However,
such observations are very challenging, in part because of strong contamination  from interstellar \ion{H}{1} emission. To our best knowledge, only five PNe have been reported 
as having a detection of the \ion{H}{1} emission line
\citep{1987A&A...176L...5T,1990ApJ...351..515T,2000RMxAA..36...51R,2002ApJ...574..179R,2006AJ....132.2566G}. Taking this into account, we 
launched a project to search for \ion{H}{1} in PNe \citep[][Paper~I hereafter]{2022ApJ...933....4O} using the new Five-hundred-meter Aperture Spherical radio Telescope \citep[FAST;][]{2011IJMPD..20..989N}, which is the most sensitive telescope at L-band frequencies of relevance here \citep{2020RAA....20...64J}.

In Paper~I we reported a FAST detection of \ion{H}{1} absorption from the young PN IC\,4997. 
This paper is devoted to investigate the dynamic interaction between PN and the interstellar medium (ISM) through the observation of \ion{H}{1} emission toward another young (600-800 year kinematic age) PN
BD+30\arcdeg3639 (BD+30 hereafter). Further details and multi-wavelength imagery, including from the 
Hubble Space Telescope ({\it HST})
 and near infrared and optical spectroscopy of BD+30 can be found in the Hong Kong/AAO/Strasbourg/H$\alpha$ PNe database
 \citep[HASH,][]{2016JPhCS.728c2008P} as catalogue entry 512.  It is known that interaction between PN ejecta and the surrounding ISM can distort the morphology of circumstellar envelopes and cause a very extensive spatial distribution of both PN ejecta and swept up materials \citep[e.g.,][]{2003ApJ...585L..49V,2010PASA...27..166S}.
Therefore, such a study might help to understand the `PN missing mass problem'.

The PN-ISM interaction has been an intensively studied topic \cite[e.g.][]{1969plne.book.....G}.
Based on a thin-shell approximation and snowplow model, \cite{1976MNRAS.175..419S} found that the central star  of interacting PN have a certain displacement with respect to nebular center. The simple analysis of \cite{1990ApJ...360..173B} showed that 
the PN-ISM interaction could lead to an asymmetrical surface brightness of PN.
\cite{1991AJ....102.1381S} suggested that the PN-ISM interaction could be summarized into
three phrases as briefly outlined below.
Initially, when nebular density is greater than that of the surrounding ISM, 
the PN is in a free-expanding phase, typically at velocities of $\sim25$--$30$\,km\,s$^{-1}$. With this nebular expansion and without further material injection, the PN density naturally decreases and eventually reaches a value comparable to that of the ambient, surrounding ISM. The upstream of the PN shell is compressed so that the surface brightness of the interacting region increases---examples include PNe like PFP1  \citep[HASH ID~2522,][]{2004PASA...21..334P}. This happens for already very evolved PN and there are few clear examples in the literature.
 As the PN density further drops, the PN shell becomes significantly distorted and the central star of the PN (CSPN) appears to be located closer to the upstream edge. Subsequently, the CSPN would move out of the PN shell, while the PN shell itself  eventually dissipates and is completely mixed with the ISM. One putative example of this is Sh 2-174  \citep[HASH ID~643,][]{1994AJ....108..978T} but this object is now classified as ionised ISM in HASH.
 The 30-year old analysis of \cite{1991AJ....102.1381S} assumed that the PN shell formed before the PN-ISM interaction.  However, observations show that an interaction may occur at various stages of PNe evolution \citep{2000ASPC..199..341R}.
 
 A more comprehensive hydrodynamic simulation of these types of interaction was performed by \cite{2007MNRAS.382.1233W}, taking into account a wide range of PN moving velocities, mass loss rates, 
 and ISM density. This work  shows that
the PN-ISM interaction can be divided into four stages: 
(i) the PN expands freely; (ii) a bow shock is formed in the interacting region, and the  density and surface brightness increase in the upstream of CSPN movement direction; (iii)
PN geometric center moves downstream with respect to CSPN; (iv) the
PN shell is completely stripped and the CSPN moves outside the PN shell. In truth there are very few examples available for study of PNe that have gone through stages (iii) and (iv).

BD+30 ($\alpha _{\rm J2000} = 19^h 34^m 45.2^s$, $\delta _{\rm J2000} = 30\arcdeg 30\arcmin 59.0\arcsec$) 
is one of the brightest young PNe in the solar neighborhood. It was used as one of four PN examples to demonstrate the value of the so-called distance mapping technique (DMT) 
based on an expansion parallax method
\citep{2020MNRAS.492.4097G}. They determined the distance to BD+30 of $D=1.3$\,kpc, while
the distance calculated from Gaia Data Release 
3 \citep{2021A&A...656A.110C} parallax is  $1.62\pm0.08$\,kpc. The Surface-Brightness radius relation statistical technique \citep{2016MNRAS.455.1459F}, arguably the best such distance technique available for the general PN population in the absence of Gaia distances, provides a distance of $D=2.22\pm0.63$\,kpc, in agreement with Gaia values  within the errors. We adopt $D=1.5$\,kpc in this paper.

The radial velocity of the PN with respect to the local standard of
rest (LSR) is $V_{\rm LSR}=-13.0\pm0.8$\,km\,s$^{-1}$ \citep{1983ApJS...52..399S}.
BD+30 is a relatively nearby, very high surface brightness PNe saturated in IPHAS (INT Photometric H$\alpha$ Survey of the Northern Galactic Plane), PanSTARRS, and $\mathit{WISE}$ (Wide-field Infrared Survey Explorer) imagery. There is clear detection of radio and extended X-ray emission \citep{2000ApJ...540..442L}. It has a Wolf-Rayet CSPN [WC9] spectrum \citep{1991A&A...252..265M} indicating close-in, fast winds.
The optical morphology of BD+30 taken from the available {\it HST} imagery is as a boxy-ellipsoid. It is classified as morphological class Eamrs according to the ERBIAS-sparm classification scheme of HASH  \citep[refer to][for details]{2022FrASS...9.5287P}.  The optical luminosity is dominated by the bright CSPN  \citep{2002AJ....123.2676L}.

H$_2$ was detected in the outer layer of the optical ring, which has
a diffuse and highly uneven spatial distribution \citep{1998ApJ...498..267S}. 
 CO is approximately symmetrically distributed at the north and south ends \citep{2000A&A...353L...5B}.
We refer the reader to \citet{2016ApJS..226...15F} for a 3D multi-wavelength reconstruction of BD+30.
\cite{1990ApJ...351..515T} reported the discovery of  \ion{H}{1} emission from BD+30 using
the Westerbork Synthesis Radio Telescope (WSRT), 
 which appears to be spatially unresolved and has a minor southeastward extension.
Among the five PNe with detected \ion{H}{1} emission, BD+30 can be reached by FAST at
the most optimum  zenith angle.
BD+30 has a moderate Galactic disk elevation, which compromises between minimizing the
contamination from the ISM \ion{H}{1} emission and maximizing the ISM density for the occurrence of substantial PN-ISM interaction.
Interferometric observations generally lack zero-spacing data, so emission sources above a certain angular size are resolved out and undetectable. This problem becomes more serious when considering issues such as the missing mass problem in the PN and the interaction between the PN and the ISM. By using a single-dish telescope with a large aperture, such as FAST, it is expected that radio emissions missed by conventional interferometry will be detected.


\section{Observation and data reduction} \label{sec:obse}

\begin{figure}[ht!]
\plotone{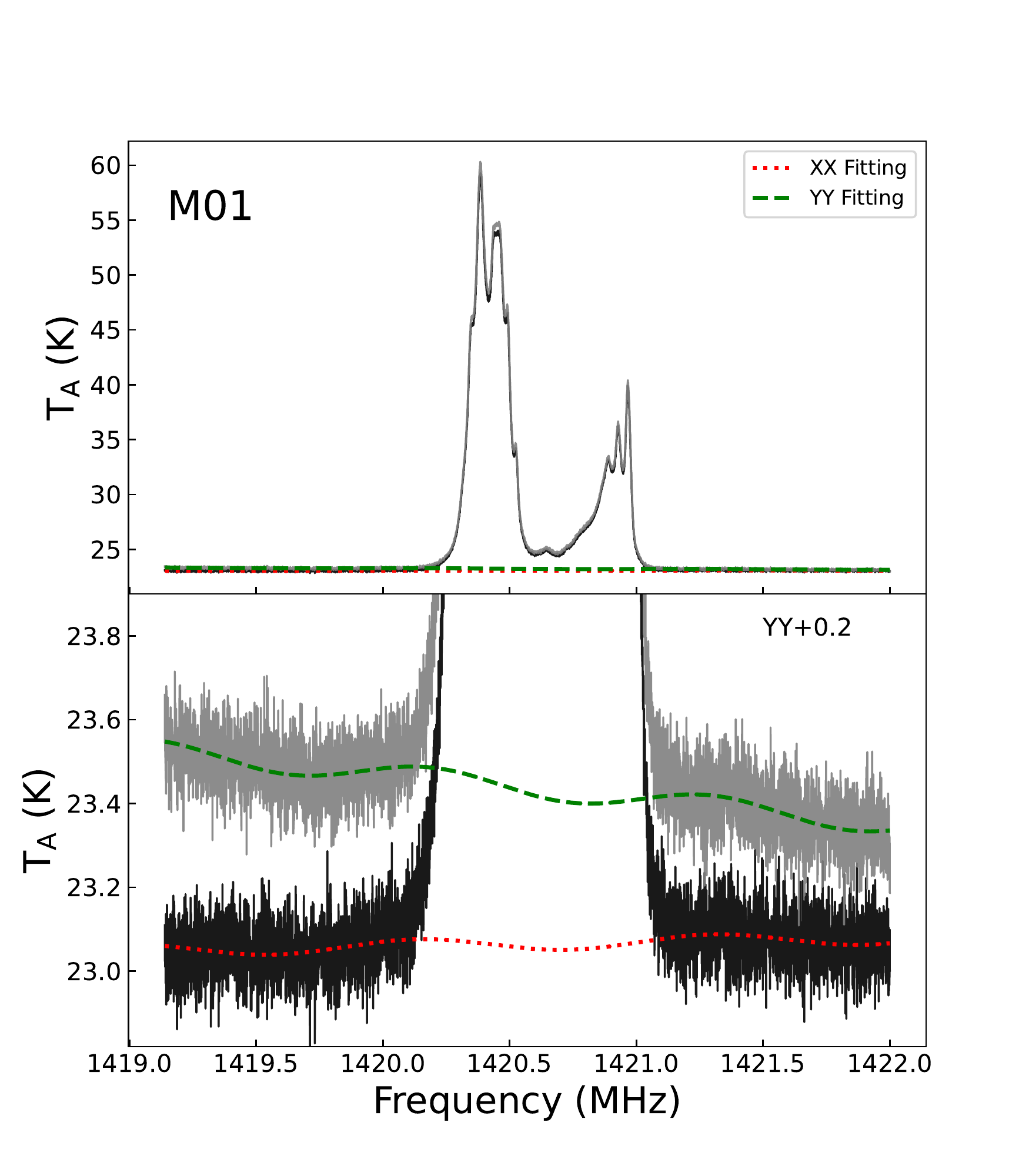}
\caption{M01 spectra of BD+30 for the two linear polarizations XX and YY as an example to show the fitting of the baseline. The lower panel is
a zoom-in of the upper panel, where the YY spectrum has been vertically shifted for a clearer view. The dotted and dashed curves represent the sinusoidal fits to the baselines of the XX and YY spectra, respectively.
 \label{fig:fitting}}
\end{figure}

The observations were performed in the tracking mode of FAST with the 19-beam
 \citep[M01--M19, see][for the arrangement of the beams]{2020RAA....20...64J}
receiver on 2021 August 9. The integration time was  40.5 minutes.
The central beam (M01), which has a half-power beamwidth (HPBW) of about 2.82$\arcmin$ at\,1420 MHz, was consistently oriented toward the target during the observation.
The standard deviation of the pointing accuracy is 7.9$\arcsec$.
The 19-beam observation mode enables us to  sample the sky over a region of
$20\arcmin\times20\arcmin$ around the target at 1.42\,GHz. The L-band (1.05--1.45\,GHz) spectra
were recorded by a backend with 1048576 channels, resulting in a frequency resolution of 476.84\,Hz and a velocity resolution of 0.01\,km\,s$^{-1}$.
For the flux calibration, a high intensity noise of about 12\,K is
injected periodically. The antenna temperature was obtained by 
\begin{equation}
T_{\rm A} = T_{\rm cal}\frac{P^{\rm cal}_{\rm on}}{P^{\rm cal}_{\rm off}-P^{\rm cal}_{\rm off}},
\end{equation}
where $T_{\rm cal}$ is the injected noise diode temperature, $P^{\rm cal}_{\rm on}$ and $P^{\rm cal}_{\rm off}$ are the power values when the diode is on and off, respectively. 
The antenna temperature was converted to flux density by $S_{\rm A} = T_{\rm A}/G$, where $G$ is the gain given by  
$G=\eta * 25.6$\,K\,Jy$^{-1}$. 
The aperture efficiency $\eta$ depends on the zenith angle at the time of the observation \citep[see][for details]{2020RAA....20...64J}.
For our observations, the $G$ values of the 19 beams  range from
 13.3 to 15.9\,K\,Jy$^{-1}$. The linear polarization channels XX and YY were separately calibrated and a baseline was subtracted.
Standing waves may be generated between the dish and the feed
cabin, complicating the baseline \citep{2020RAA....20...64J}.
Therefore, we used a sinusoidal function to fit the baseline.
Nevertheless,  the standing wave has a low amplitude
(see Figure \ref{fig:fitting} for an example), 
and barely affects our results.
The XX and YY spectra then were co-added to obtain the final spectra.  
Based on the line-free spectral regions, we obtained a root mean square noise of the final spectrum of 1.05\,mJy.

\section{Analysis and results} \label{sec:resu}

Figure \ref{fig:fit01-19} displays the spectra of the 19 beams. The overall spectral 
shapes are quite similar among the different beams, suggesting that the \ion{H}{1} spectra are dominated by interstellar emission. The observed lines are clearly grouped into two main velocity ranges,
$-110$ to $-50$\,km\,s$^{-1}$ and $-40$ to $+40$\,km\,s$^{-1}$. The first negative velocity group (in terms of LSR) has a sharp blue wing and
a broad red tail down to low flux levels, while the latter is much stronger and does not exhibit a broad red tail. Both major velocity groups are composed of several sub-peaks with different relative intensities for
different beams.

\subsection{Tentative detection of H~I absorption}

\begin{figure}[ht!]
\plotone{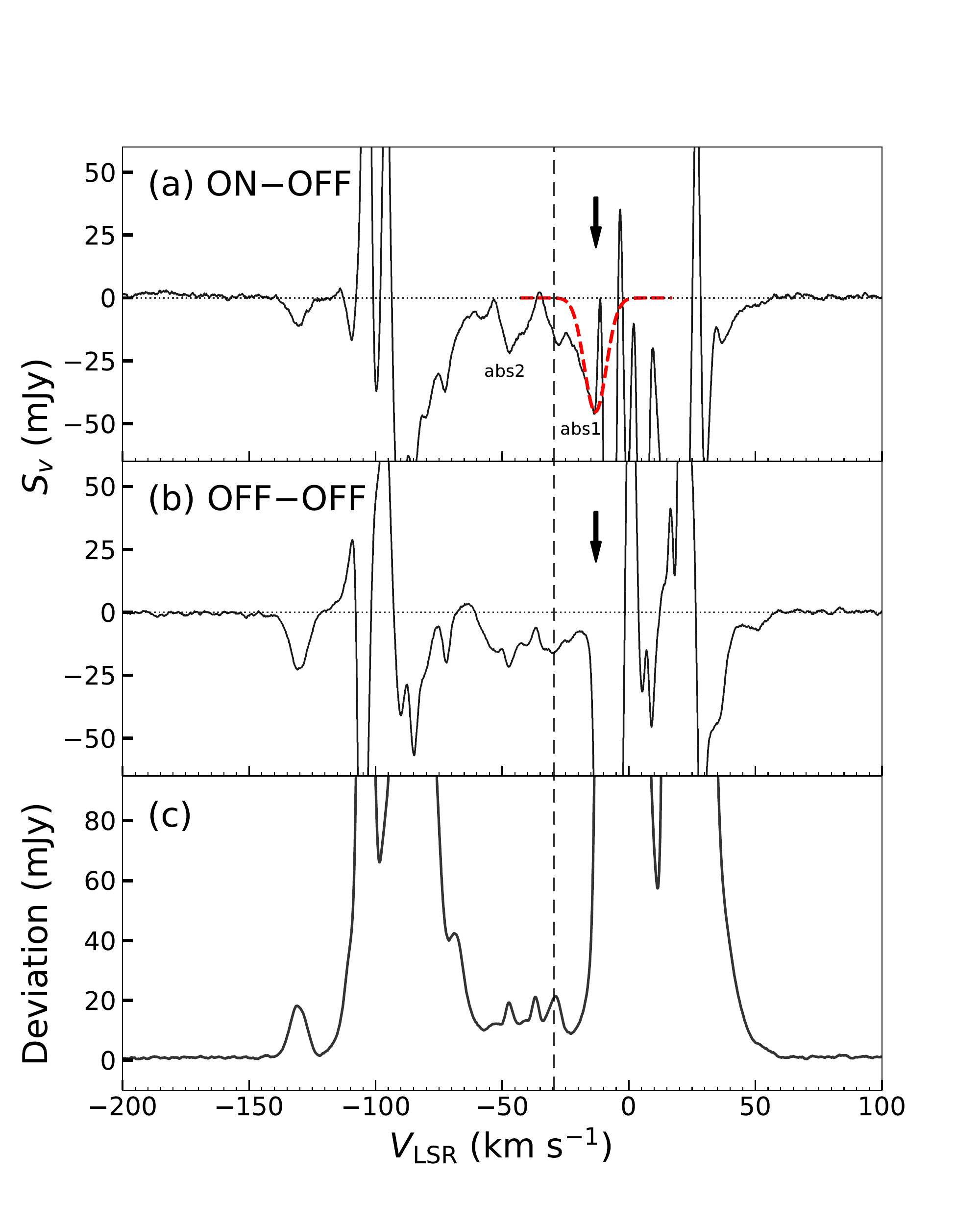}
\caption{(a) The on$-$off spectrum obtained from the M01 spectrum subtracted
by the average spectrum of M02--M07.
(b) The off$-$off spectrum obtained from the average of the west and east spectra (M02 and M05) subtracted by the average of the north and south spectra (M03, M04, M06, and M07).
(c) The standard deviation of the off spectra.
 The arrows mark the systemic velocity of the PN.
Two tentatively detected  \ion{H}{1} absorption features (denoted by `abs1' and `abs2') are reported here.
 The red dashed curve is a Gaussian fit to `abs1'.
 The vertical dashed line indicates the velocity
 of the \ion{H}{1} absorption reported
  by \cite{1990ApJ...351..515T}.
 \label{fig:spc}}
\end{figure}

Following Paper I, we take the sky sampled by the beams surrounding M01 as the
off-source comparison positions. Therefore, the off-source spectrum is obtained by
averaging the M02--M07 spectra and is subtracted from the M01 spectrum to
obtain the on$-$off spectrum, as shown in Figure \ref{fig:spc}a.
For a null check, we obtain two off-position spectra by averaging 
the spectra observed by the west and east beams and that by the north and south beams.
The off$-$off spectrum is shown in  Figure \ref{fig:spc}b.
\cite{1990ApJ...351..515T} have reported a detection of
\ion{H}{1} absorption at $-29.5$\,km\,s$^{-1}$. That feature is clearly shown in our on$-$off spectrum. 
Apart from that, the on$-$off spectrum exhibits
a broad dip over the velocity range from $-40$ to $-50$\,km\,s$^{-1}$.
However, as they are also seen in the off$-$off spectrum,
we cannot exclude the possibility of spurious features caused by the subtraction of the off spectrum.
The on$-$off spectrum  exhibits another deeper dip at $-13$\,km\,s$^{-1}$, near the systemic velocity of the PN, which is not seen in the off$-$off spectrum. 
In Paper I, we detected such a feature in the young PN IC\,4997
and attributed it to \ion{H}{1} absorption stemming from 
the front side of an expanding circumstellar shell.
The \ion{H}{1} absorption is against the free-free continuum of the
relatively compact ionized region, and thus is well confined in the central
beam only. In order to validate the detection of \ion{H}{1} absorption in BD+30, we
compute the deviation of the M02-M07 with respect to their average, as
shown in  Figure \ref{fig:spc}c. At the velocity of the dip, the deviation
is much larger than the depth of the dip, indicating a large variation between
different off-position spectra (relative to the circumstellar absorption). The subtraction of the background 
emission is the dominant uncertainty.
Therefore, we cannot definitively rule out the possibility
that the dip seen in Figure \ref{fig:spc}a is a spurious absorption.
Nevertheless, we note that such a dip is also visible in the BD+30 spectrum shown in
 \cite{1990ApJ...351..515T}. 
 
Based on these current data, we consider the detection of
 circumstellar \ion{H}{1} absorption in BD+30 as tentative.
Our convincing detection of  \ion{H}{1} absorption in Paper I for PN IC\,4997 can be partly attributed to the considerable difference between the velocities of
IC\,4997 and interstellar \ion{H}{1} lines. For BD+30 however, such discrimination is
hindered  due to the closer velocities between the PN and the observed interstellar 
\ion{H}{1} lines.

A Gaussian fit is performed to derive the integration strength and
width of the absorption feature near the systemic velocity
of the PN.
The absorption line shows a sharp red edge (Figure~\ref{fig:spc}a), where a 
\ion{H}{1} emission line, as detected by \citet{1990ApJ...351..515T}, contributes to the flux. 
To this end, we fix the velocity of
the dip minimum as the central velocity ($V_{\rm m}=-13.29\pm0.26$\,km\,s$^{-1}$)
and consider the blue side only for the fitting. It results in 
a full width at half maximum (FWHM) of $10.45\pm0.59$\,km\,s$^{-1}$, a maximum absorption flux density of $-45.3\pm2.5$\,mJy, and an integrated flux density of $-630\pm62$\,mJy\,km\,s$^{-1}$.
Following \citet{1990ApJ...351..515T} and assuming a
continuum flux density at 21\,cm of  $238\pm15$\,mJy, we obtain
the integral of the optical depth ($\tau_v$) over velocity to be
 $\int\tau_v dv=2.65\pm0.09$\,km\,s$^{-1}$.

\subsection{Extended H~I emission}

\begin{figure}[ht!]
\plotone{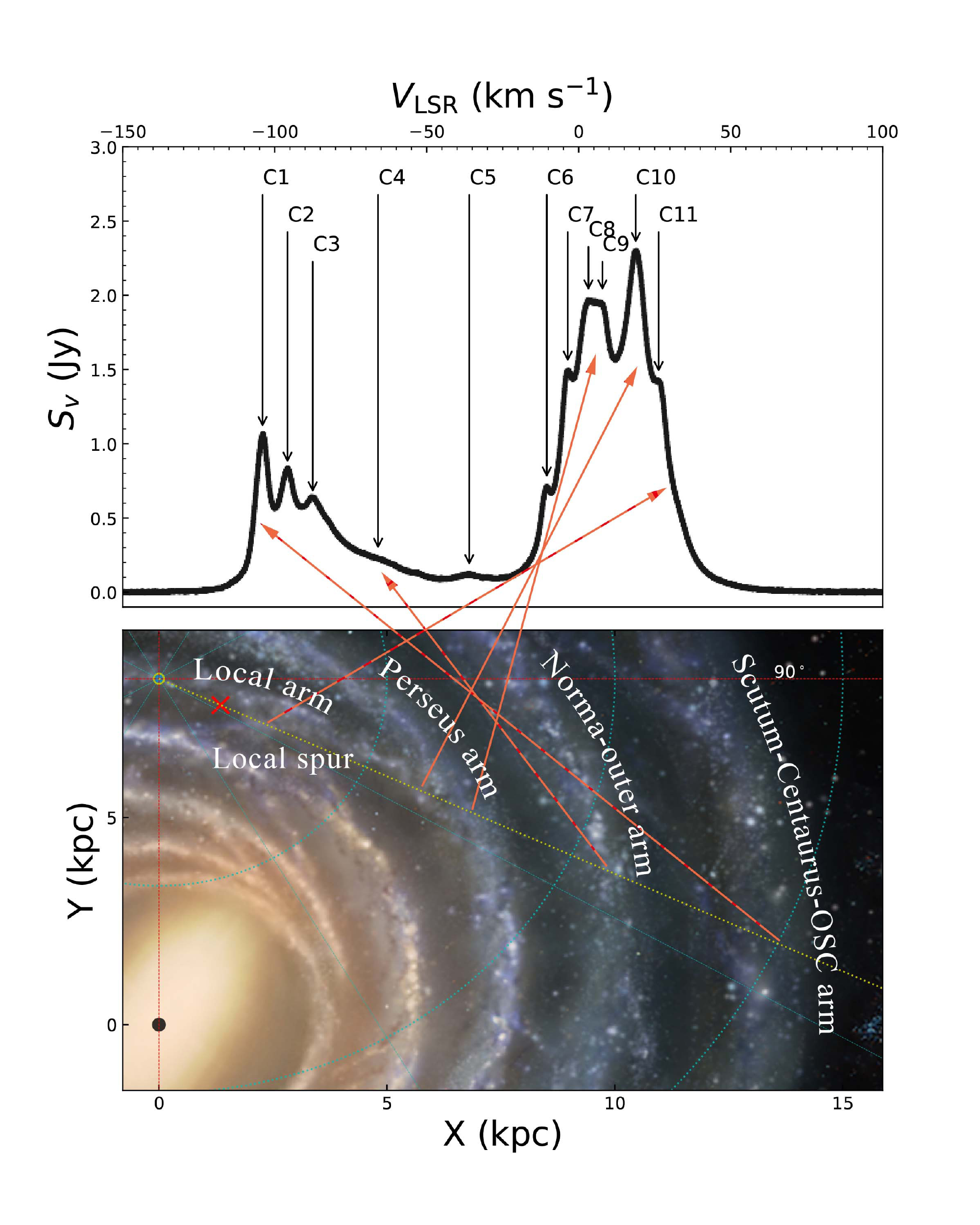}
\caption{The association between the spectral components and
the Galactic spiral arms identified by their velocities, as specified by the red arrows.
The upper panel shows the M01 spectrum, where 
C6 and C7 correspond to the unsolved circumstellar feature
in the spectrum of \cite{1990ApJ...351..515T}. 
The lower panel displays the Galactic spiral arms
(image credit: Xing-Wu Zheng \& Mark Reid  BeSSeL/NJU/CFA),
where the open circle is the position of the Sun, the filled circle is the Galactic center, the red cross is the position of BD+30, and the dotted yellow line represents the line of sight.
 \label{fig:comarm}}
\end{figure}

If the atomic shell is very extended, \ion{H}{1} can be observed in emission against the Galactic
synchrotron background.
 The \ion{H}{1} emission line detected by \cite{1990ApJ...351..515T}
lies in the velocity range of $-11$ to 2\,km\,s$^{-1}$ and peaks on the
red side.  The M01 spectrum clearly exhibits two peaks at the same 
velocity range, and the peak on the red side is stronger  (see Figure~\ref{fig:comarm}). Because the spectral 
resolution of \cite{1990ApJ...351..515T} is 4.12\,km\,s$^{-1}$, much poorer than ours, 
we can conclude that the emission feature discovered by them  
actually consists of two lines that are well resolved by the FAST observations.

\begin{figure}[ht!]
\plotone{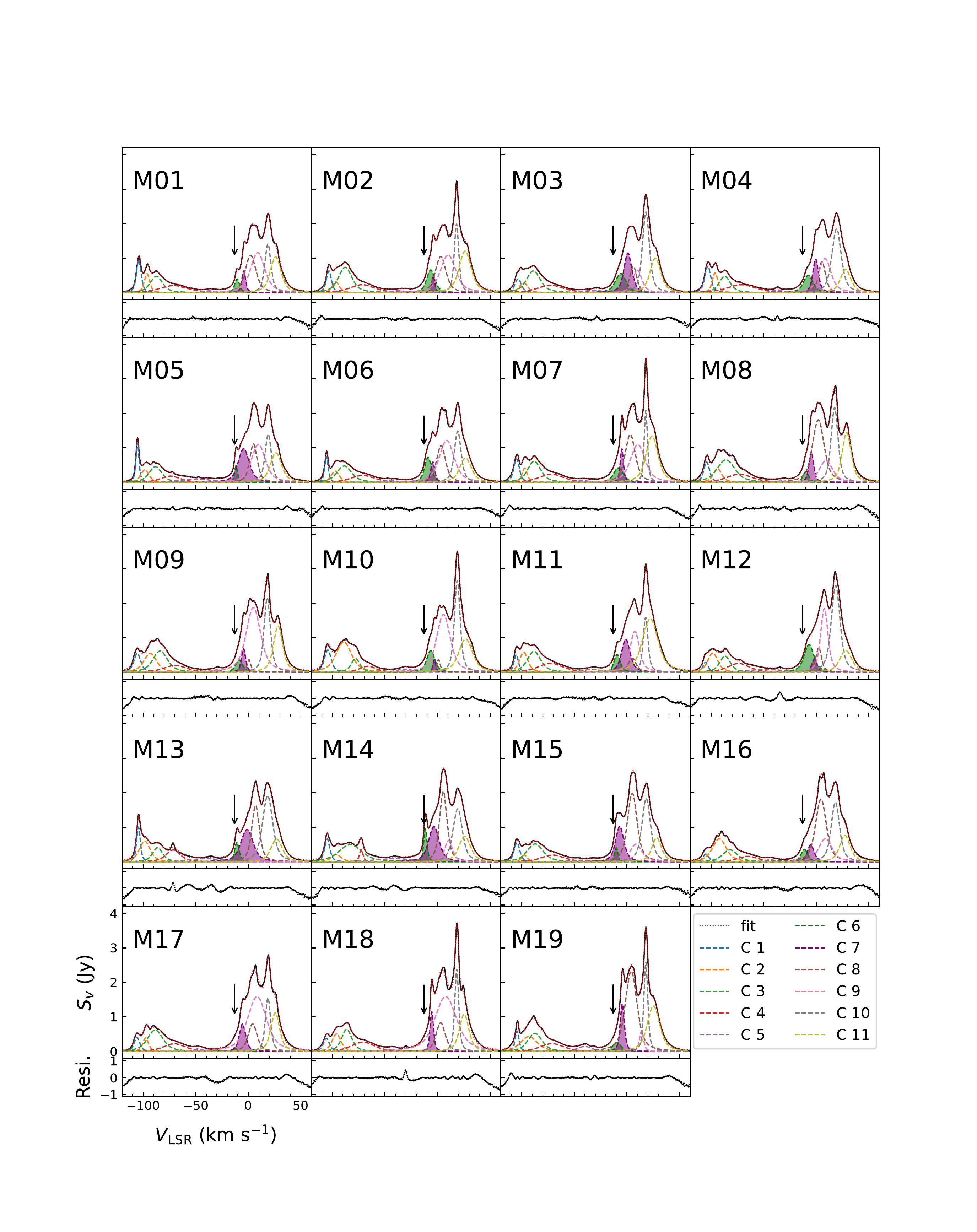}
  \caption{The spectra obtained with the 19 beams (M01--M19). The observed
  and fitted spectra are shown in black and red, respectively.
 The colorful dashed curves represent  different spectral components (C1--C11). 
     The arrows indicate the systemic velocity of BD+30.
The green and purple shallows correspond to the  unsolved \ion{H}{1} feature detected by \cite{1990ApJ...351..515T}.
 The fitting residuals with respect to $S_\nu$ are displayed in the lower part of each panel.
  \label{fig:fit01-19}}
\end{figure}

In order to derive the strengths and widths of the two emission lines that
are potentially associated with BD+30, we
perform a spectral decomposition. By carefully examining the M01--M19 spectra, we find that at least 11 
components (C1--C11) are required to reproduce these spectra, as indicated in Figure~\ref{fig:fit01-19} 
(also see Figure~\ref{fig:comarm} for a zoom-in
of the M01 spectrum). C6 and C7 correspond to the \ion{H}{1} emission 
that is claimed to originate from BD+30 by  \cite{1990ApJ...351..515T}. 
Based on the current  Galactic rotation model \citep{2014ApJ...783..130R,2016ApJ...823...77R}, 
we can infer the Galactic positions emitting the interstellar spectral components,
as illustrated in Figure~\ref{fig:comarm}.
The negative-velocity group, C1--C5, stems from high-velocity halo gas \citep{2015MNRAS.449..220M} or the 
spiral arms at a distance of about 8.7--20.5\,kpc, which does not affect the
extraction of C6 and C7. 
However, C9--C11 are badly blended with C6 and C7. 
The Galactic rotation model indicates
that the distances to the regions emitting C11, C10, and C8--C9 are
about 2.2, 4.9 and 6.6\,kpc, respectively, corresponding to
the Local spur (longitude range 50$^{\circ}$--78$^{\circ}$ and velocity range 4--49\,km\,s$^{-1}$, 
\citealt{2016ApJ...823...77R}), the tail of the Local arm (longitude range 55$^{\circ}$--282$^{\circ}$ and 
velocity range $-23$--25\,km\,s$^{-1}$, \citealt{2013ApJ...769...15X}), and the Perseus arm (longitude range 
25$^{\circ}$--270$^{\circ}$ and velocity range $-48$--28\,km\,s$^{-1}$, \citealt{2016ApJ...823...77R})  along
the line of sight to BD+30. C11 is the tangent-point emission.

The velocities of C6 and C7 do not coincide with those of spiral arms. 
The properties of individual \ion{H}{1} lines are 
measured by a multiple Voigt profile fitting procedure,
for which we use the Python LMFIT package
\citep[nonlinear least-square minimization and curve-fitting;][]{2021zndo...5570790N}. The fitting results 
are shown in Figure~\ref{fig:fit01-19}, where C6 and C7 are highlighted. The residual errors of fitting for 
C6 and C7 are better than 14\% of the peak brightness.
Table \ref{tab:fitting result} presents the 
integrated flux density over velocity ($\int S_v dv$),
the peak LSR velocities 
($V_{\rm p}$), the FWHMs, and the peak flux densities ($S_{\rm p}$)
of C6 and C7 derived from the spectral decomposition.
These values vary spatially, and as a result, the on$-$off spectrum generates two narrow
spikes redward from the absorption in Figure \ref{fig:spc}a.

For the M01 spectrum, 
 C7 lies on the red side of C6 and appears to have
 a larger $S_{\rm p}$,
while the velocity of C6 ($-10.63$\,km\,s$^{-1}$)
is closer to the systemic velocity of the BD+30 than that of C7 
($-4.33$\,km\,s$^{-1}$). This is entirely 
consistent with the lower spectral resolution observation of 
\cite{1990ApJ...351..515T}, where the \ion{H}{1} emission line
lies redward of the systemic velocity of the PN and has
an enhanced red shoulder.
On the other hand, the  observation of \cite{1990ApJ...351..515T}
with a synthesized beam size of $11.5\arcsec\times23\arcsec$
shows that the \ion{H}{1} emission is spatially unresolved,
and thus should be exclusively confined in the M01 beam.
However, C6 and C7 are exhibited in all M01--M19 spectra
(See Figure~\ref{fig:fit01-19} and Table \ref{tab:fitting result}),
suggesting that the \ion{H}{1} emission has an extent of
at least $15\arcmin$.  It is clear that the extended structure
has been filtered out in the interferometric observation
due to the lack of antenna pairs with small separations, which can be well 
compensated by the FAST 
observations. C6 and C7 can be tentatively attributed to the front and back 
sides of a large expanding shell. We  assume that their velocities are not symmetric
with respect to $V_{\rm LSR}$ because of an interaction between the PN and the ISM.

Under the optically thin assumption, the mass of \ion{H}{1}
can be determined  by
\begin{equation}\label{m2}
M_{\rm HI} = 2.36\times10^{-4}\left( \frac{D}{\rm kpc} \right)^2\left( 
\frac{\int S_v dv}{\rm mJy\,km\,s^{-1}} \right){\rm M}_\sun.
\end{equation}
Adding the integrated intensities
detected in M01-19 spectra, we derive the total mass of \ion{H}{1} in C6 and C7 
to be 52.3 and 79.2\,${\rm M}_\sun$, respectively.
Those are the lower limits of the mass of the postulated shell 
because of the existence of the unsampled regions between beams.
The large mass allows us to rule out the possibility of C6
and C7 as ejected stellar materials. A natural conjecture is
that C6 and C7 are the interstellar gas swept-up by the stellar wind emanating from BD+30. 
Nevertheless, the swept ISM mass is formidably large for a PN.
In the models of \cite{2007MNRAS.382.1233W},
the swept-up shell stalls when the accumulated mass
reaches about ten times the ejecta mass, a value
smaller than that of C6 and C7. 
Therefore, we cannot completely 
 exclude the possibility that C6 and C7 originate from 
 clouds overlapping along the
 line of sight of BD+30 with coincidentally similar velocities.
If this were the case, our observations could disprove the conclusion
of \cite{1990ApJ...351..515T}, and BD+30 should be omitted from
the list of PNe with detected circumstellar \ion{H}{1} emission.

\section{Discussion} \label{sec:disc}

Assuming that our tentative detection of the \ion{H}{1} 
absorption (Fig. \ref{fig:spc}c) is real and using the same method described
in Paper I, we can estimate the
expansion velocity, mass, and column density of
the neutral shell traced by the absorption at
$-13.29$\,km\,s$^{-1}$.
The expansion velocity along the line of sight 
is derived from 
$V_{\rm LRS}-V_{\rm m}+{\rm FWHM}/2=5.5$\,km\,s$^{-1}$.
An optical spectroscopic study has shown that the N$^+$ and O$^{++}$
ionized regions have an expansion velocity of 28\,km\,s$^{-1}$ and 
35.5\,km\,s$^{-1}$, respectively \citep{1999MNRAS.309..731B}. It is well known that BD+30 contains
a central hot-bubble \citep[e.g.][]{2018A&A...620A..98H} also revealed as 
extended X-ray emission \citep{2000ApJ...540..442L}.
The ionization and shock heating may results in a radial pressure gradient, 
causing an acceleration of the ionized gas.
Another potential cause for the low \ion{H}{1} velocity is
that the proposed interaction with the ISM has caused
significant deceleration of a much more extended neutral shell, indicating the presence of a stalled astrosphere.
 Under a thin-shell assumption, 
the angular radius of the \ion{H}{1} envelope
is about equal to that of the ionized region \citep[$4^{\prime \prime}$.93,][]{2021MNRAS.503.2887B}. The excitation temperature of \ion{H}{1} in the PN
is poorly known, but it is reasonable to assume an intermediate value of 
 100\,K \citep[see][and Paper I for a discussion]{1986ApJ...305L..85A}. 
From the absorption feature at $-13.29$\,km\,s$^{-1}$,
the column density and  mass of the \ion{H}{1}
are thus determined 
to be N$_{\rm abs1} = 4.8 \times 10^{20}$ cm$^{-2}$ and
M$_{\rm abs1}=3.1 \times 10^{-2}$ M$_\sun$, respectively.
From the continuum flux density we estimate the 
 mass of ionized gas to be $3.7 \times 10^{-2}$ M$_\sun$.
Consequently, the mass ratio of atomic to ionized gas is about 0.8,
which is close to the value in another young PN IC 4997 (Paper I). It should be emphasized that 
the atomic/ionized gas mass ratio is independent of
the distance and nebular size under the thin-shell assumption.
The number of   neutral hydrogen atoms  in BD+30 is estimated as $4\times 10^{55}$.

\begin{figure}[ht!]
\plotone{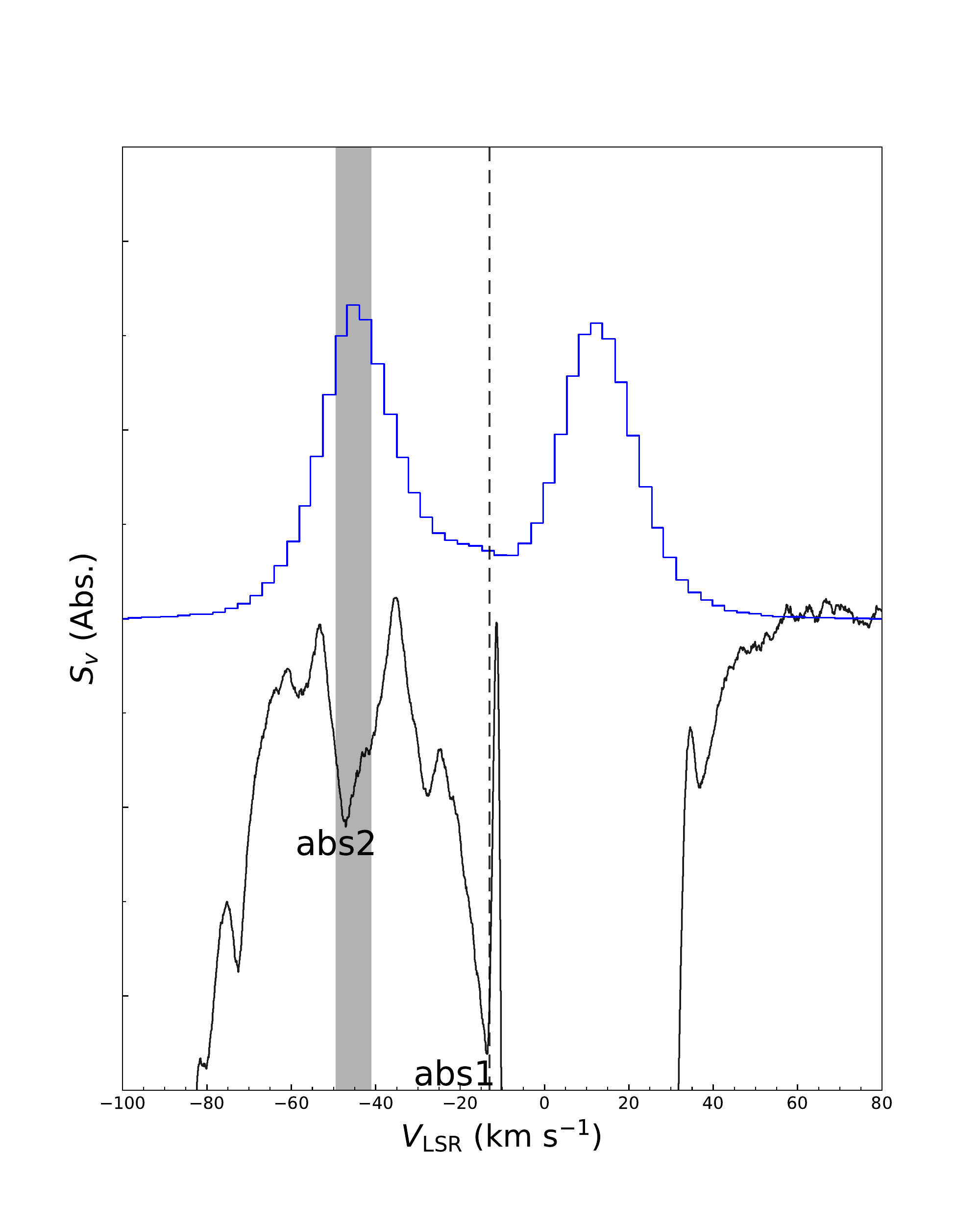}
\caption{Comparison of the \ion{H}{1} absorption
(lower) and the [\ion{N}{2}]$\lambda6548$ emission line (upper) taken from \citet{1999MNRAS.309..731B}.
The shaded area represents the velocity of
the blue peak of  [\ion{N}{2}]$\lambda6548$. The dashed line marks the systemic velocity of the PN. Note that the spectral ranges with strong contamination from interstellar emission have been blanked out.
\label{fig:com}}
\end{figure}

Figure~\ref{fig:com} presents a comparison between our spectrum and the [\ion{N}{2}]$\lambda6548$ emission line observed by \citet{1999MNRAS.309..731B}.
The [\ion{N}{2}] line traces the velocity of  the outermost ionized gas, which can be 
accelerated by the pressure from the ionized gas.
It exhibits two peaks arising from
the front and back sides of the expanding ionized shell; the blue one has a velocity coinciding with the  dip seen in our \ion{H}{1} spectrum between 
$-40$ and $-50$\,km\,s$^{-1}$. This leads us to the speculation that this dip is associated
a \ion{H}{1} absorption tracing the outer boundary of the ionized nebula, although the identification
is not certain because the velocity is unfortunately similar to that of the interstellar \ion{H}{1}. 
A further support for this speculation comes from
the wide-aperture observation of \citet{2000A&A...358..321N}
who show that [\ion{O}{1}] and [\ion{S}{2}] emission lines have a similar width
suggesting that the velocity remains similar across the ionization front.
However, even if validated,
the absorption over the velocity range from
 $-40$ to $-50$\,km\,s$^{-1}$ 
has an  equivalent width comparable to 
that of the absorption at $-13.29$\,km\,s$^{-1}$, 
suggesting the mass of the absorption gas to be M$_{\rm abs2}\sim 10^{-2}$ M$_\sun$, still far less
than the missing PN mass.

According to \citet{1995MNRAS.276.1101A}, BD+30 has a very low He/H abundance
ratio  ($\sim 0.02$). Even correcting for helium abundance, the total gas mass is still fairly small compared to the upper theoretical limit of PN mass.
 Therefore, the observation of \ion{H}{1} absorption as in BD+30 cannot solve the `PN missing mass problem'.  The uncertainties of the mass estimation primarily stem
 from the assumed radius of the atomic shell.
Nevertheless, the observations of \citet{2000A&A...353L...5B} demonstrate that the 
molecular and ionized regions of BD+30 have a similar size, which 
justifies the thin-shell assumption for the calculation of  the \ion{H}{1} mass.


Is the missing mass contained in extended nebulosity that has been
stripped by the ram pressure of the ISM? To settle this question we
need to investigate the interaction between BD+30 and the ISM.
BD+30 has a proper motion of $-2.3\pm0.03$\,mas\,yr$^{-1}$ and $-8.8\pm 
0.03$\,mas\,yr$^{-1}$ in right ascension and declination, respectively
\citep{2021A&A...649A...1G}. The values are in good
agreement with those given by \cite{2008A&A...479..155K}
but with lower uncertainties.
This corresponds to a velocity of 
$14.1$\,km\,s$^{-1}$ eastward and $62.6$\,km\,s$^{-1}$ northward
at an adopted distance of 1.5\,kpc, and thus the position angle
of the proper motion is 192$^{\circ}$.7 east of north.
The heliocentric velocity of BD+30 is $-$31.4\,km\,s$^{-1}$
\citep{1983ApJS...52..399S}.
Assuming that the ISM follows the Galactic rotation with no peculiar motion, we can calculate
the velocity of BD+30 relative to its ambient ISM
($V_*$). Using the Oort constants of
$A=15.3$\,km\,s$^{-1}$\,kpc$^{-1}$ and $B=-11.9$\,km\,s$^{-1}$\,kpc$^{-1}$, we obtain the radial and tangential velocities of the local ISM with respect to the position of the sun to be $V_r=17.7$\,km\,s$^{-1}$ and $V_t=-32.4$\,km\,s$^{-1}$, respectively.
The ISM velocities can be converted to
heliocentric values by adopting a solar motion with respect to the LSR of
$(U, V, W)_\odot= (10.6, 10.7, 7.6)$\,km\,s$^{-1}$ 
\citep{2019ApJ...885..131R}, where the velocity components 
$U$, $V$, and $W$ are respectively defined along the radial, orbital, and
vertical directions of the Galactic plane.
We finally have a $V_*$ value of $\sim50$\,km\,s$^{-1}$.
The Galactic disk elevation of BD+30 is about 130\,pc, where
the ISM is largely composed of cold neutral gas
\citep{2003ApJ...586.1067H}.
The large motion velocity and the high ISM density provide favorable
conditions for the interaction between stellar wind and ISM, even
for such a young PN. However, there is scant evidence in the optical for any obvious ISM-PN interaction probably
because of insufficient ultraviolet radiation to excite optical lines in extended regions.

\begin{figure}[ht!]
\plotone{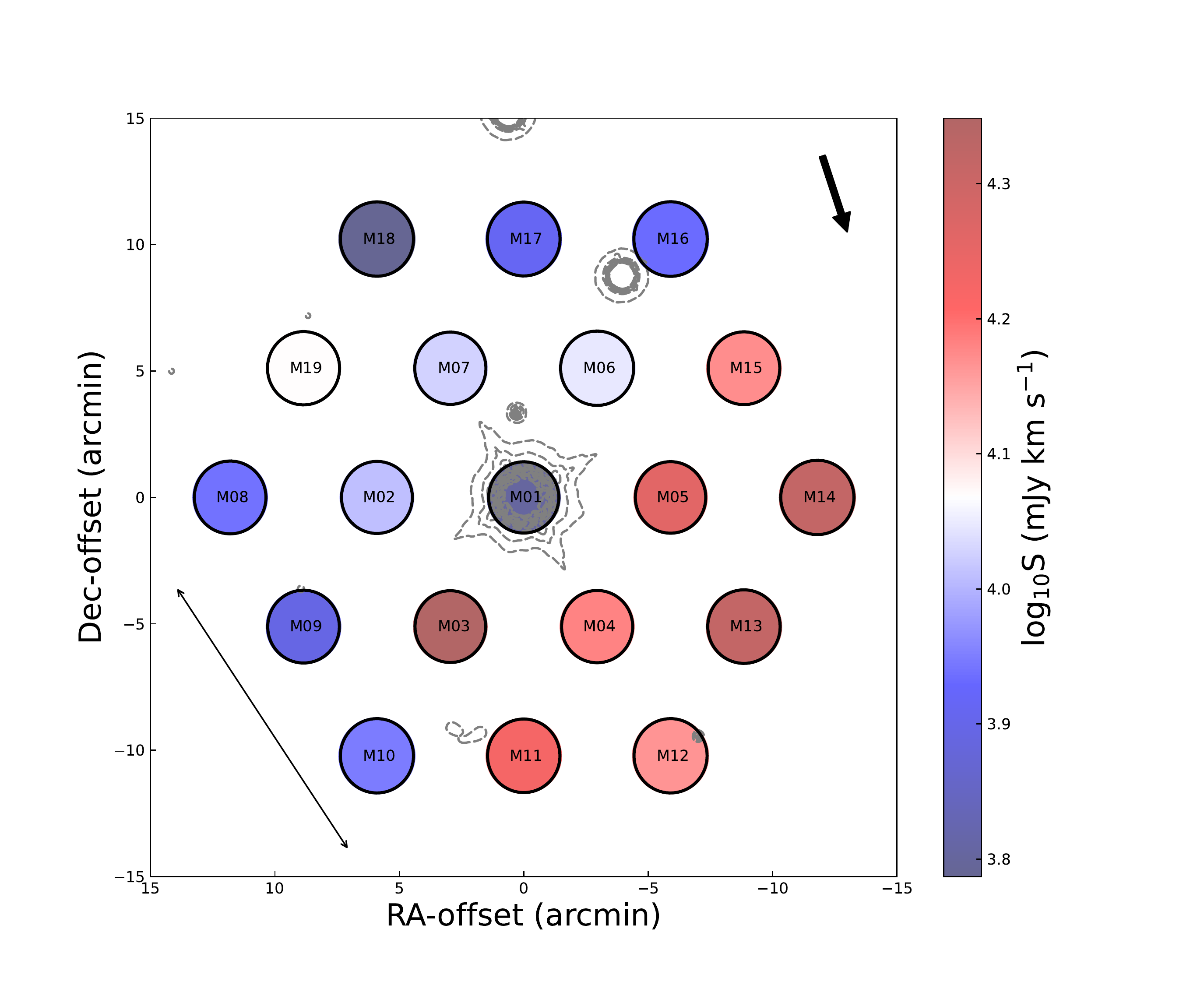}
\caption{Spatial distribution of the integrated intensities of 
C6 and C7. The arrow in the upper right corner 
indicates the direction of proper motion of BD+30. The direction of the Galactic disk is denoted by a double headed arrow in the lower left corner.  The background is the contour map of the W4 band of the $\mathit{WISE}$.
\label{fig:intsum}}
\end{figure}

The statistical study of \citet{2012A&A...541A..98A} shows that
PNe lying close to the Galactic plane are most likely to interact
with molecular and cold neutral clouds.
\citet{2015MNRAS.449..220M} have demonstrated that 
a \ion{H}{1} observation with
a single-dish telescope is more sensitive to detect
the diffuse low surface brightness emission surrounding the
carbon-rich asymptotic giant branch (AGB) star IRC+10216 compared to the Very 
Large Array. A  \ion{H}{1} emission tail formed by
the interaction with the ISM has been discovered
in an AGB Mira variable \citep{2008ApJ...684..603M}.
It has been shown that the asymmetric distribution of 
\ion{H}{1} emission, the increase of flux in the interaction region, and the 
formation of bow shock can be perceived as a sign of wind-ISM
interaction \citep{1996ApJS..107..255T,1998ApJ...495..337D}.
In Figure~\ref{fig:intsum} we examine the spatial distribution
of the integrated intensity of C6 and C7. It is shown that
the outer beams detect stronger \ion{H}{1} emission than the central
beam. The \ion{H}{1} emission is obviously enhanced toward the southwest, which 
is roughly aligned with the direction of the proper motion.
This is a typical characteristic of the interaction between a moving
PN and the ISM that causes the accumulation of gas upstream of the direction of 
PN motion.
Apart from that, the density gradient of the ISM could also affect the 
morphology of the
circumstellar envelope, and possibly lead to an enhancement of \ion{H}{1} emission
toward the Galactic disk. This is not supported by our observations (see Figure~\ref{fig:intsum}) partly because
 the density scale height is much larger than the nebular size.
 
How much is the ISM gas that could be swept up by the motion and expansion of BD+30?  Assuming that the stellar winds were launched during the AGB phase 
and lasted $10^5$ years and taking the expansion velocity 
to be a typical value $V_{\rm exp}=20$\,km\,s$^{-1}$, 
we find that
these earlier stellar winds, prior to final PN envelope ejection, can reach up 
to a radius of 2\,pc in the head-on direction, corresponding to
an angular radius of $\sim5$\,$\arcmin$. Considering the significant uncertainty of
such a crude estimation, this value is not far from the observed extent of \ion{H}{1} emission (Figure~\ref{fig:intsum}).
If BD+30 crosses the ISM at a constant velocity of 
$V_*=50$\,km\,s$^{-1}$, its central star has travelled a length of about 5\,pc 
since it joined the AGB phase. 
A runaway shell expanding at a constant velocity ($V_{\rm exp}<V_*$)
would take on a raindrop-like shape. 
A simple geometric calculation shows that
the volume of the plowed ISM gas is about 42\,pc$^3$.
The cold neutral medium has 
a density of $n_{\rm H}=10$--100\,cm$^3$ \citep{2004come.book...33W}, and
 thus the mass of the swept ISM gas is estimated to be
10--100\,${\rm M}_\sun$. If the central star of BD+30 has been decelerated by dynamic friction
(i.e. the initial velocity is larger than 50\,km\,s$^{-1}$),
the mass of the swept ISM gas will be larger
than the above estimated value. Therefore,
the mass value estimated for the swept material is
compatible with that derived from the intensities of the \ion{H}{1} emission lines ($\sim100$\,M$_\sun$). Despite the severe uncertainties in these estimates, there is a reasonable agreement that supports the swept-material interpretation.
It should also be noted that the flow surrounding the PN-ISM interface could constrain the nebular expansion and
cause a  smaller mass of the swept ISM gas. 
A sophisticated modelling is required to obtain an accurate estimation, which is beyond the scope of this paper.
According to \citet{1990ApJ...360..173B}, the PN-ISM interaction
is observable when the nebular density decreases to a critical density,
 $n_{\rm crit} = [(V_*+V_{\rm exp})/{c_s}]^2n_{\rm H}$, where 
$c_s$ is the isothermal sound speed. Assuming
$n_{\rm H}=10$\,cm$^{-3}$ and $c_s=10$\,km\,s$^{-1}$, we obtain
$n_{\rm crit}=490$\,cm$^{-3}$. The density of the \ion{H}{1} shell 
is presumably lower than $n_{\rm crit}$, and thus could ensure the occurrence
of the PN-ISM interaction.

\begin{figure}[ht!]
\plotone{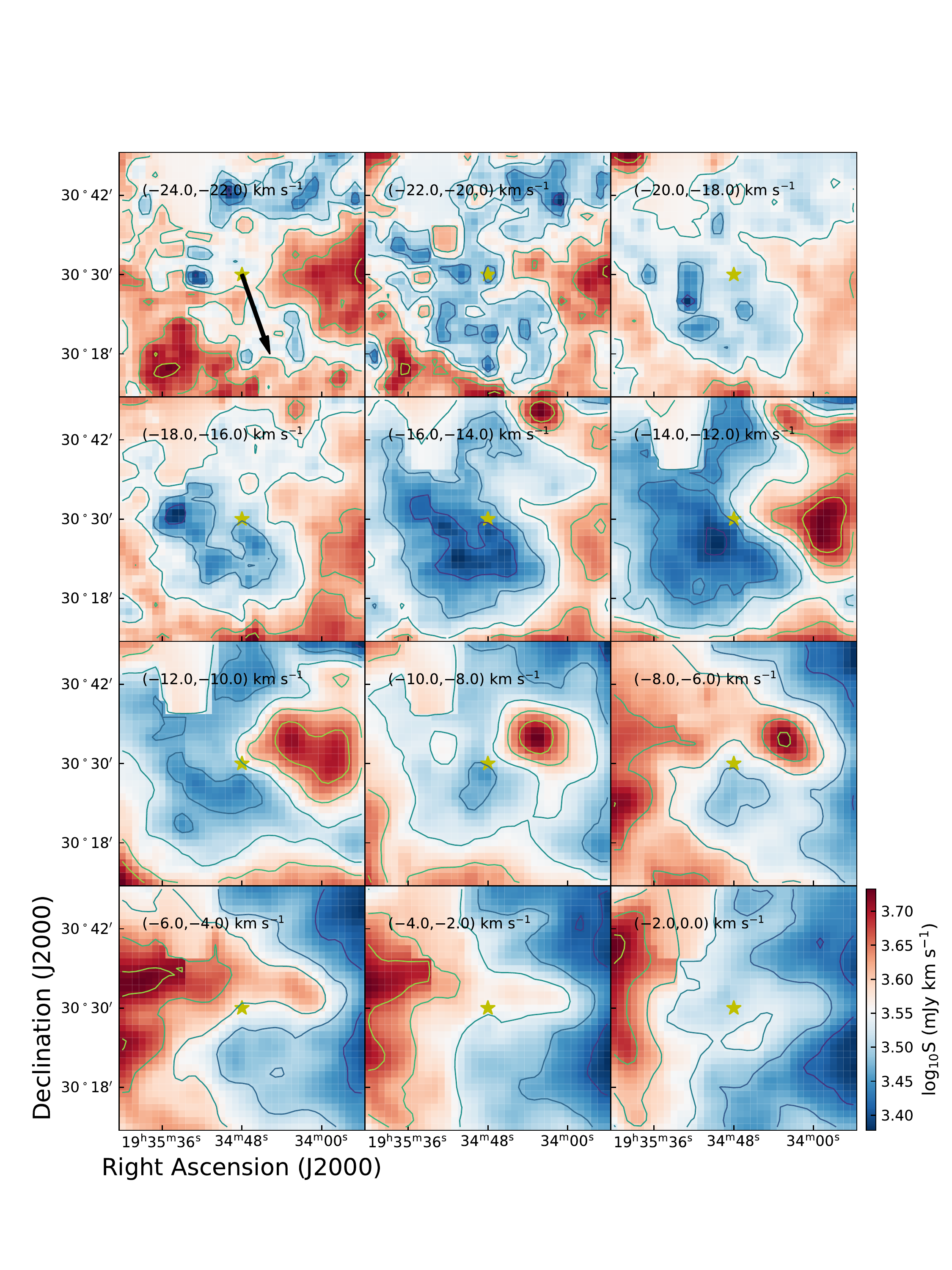}
\caption{Channel maps from the GALFA-HI survey
in the velocity range from $-24$ to 0\,km\,s$^{-1}$ with a channel width of 2\,km\,s$^{-1}$. 
The yellow star marks the position of BD+30.
The arrow in the upper left panel indicates the direction of the proper motion.
 \label{fig:cm}}
\end{figure}

Also shown in Figure~\ref{fig:intsum} is the contour map of the {\it WISE} W4 (22\,$\mu$m) band. This appears 
unresolved, as seen in the {\it WISE} images in HASH  for 
convenience.
It has been shown that
thermal emission  from cold dust in the mid-infrared is a good tracer for
extended PN-ISM structures \citep[e.g.][]
{2012ApJ...755...53Z,2018MNRAS.475..932R}.
However, we do not see any dust emission associated with the \ion{H}{1} shell.
To further validate our observations, we extracted the data of the Galactic 
Arecibo L-Band Feed Array \ion{H}{1} (GALFA-\ion{H}{1}) survey 
\citep{2011ApJS..194...20P,2018ApJS..234....2P}. Although the GALFA-\ion{H}{1}
data have a lower sensitivity than the FAST observations, they cover a 
contiguous sky area surrounding BD+30. The channel maps of the \ion{H}{1}
emission is shown in Figure~\ref{fig:cm}. We can see that the maps in
the velocity range of
$-18$ to $-8$\,km\,s$^{-1}$ exhibit a cavity in the position of BD+30 and
the emission appears to be stronger in the leading edge of the PN.
This is consistent with our observations and may be considered as 
a possible trail that has be plowed by a moving expending shell in the ISM. 
Due to the large velocity shear at the PN-ISM interface, a Kelvin-Helmholtz instability is developed and 
induces turbulent mixing in the wake.
If our interpretation is correct, it would mean that the \ion{H}{1} shell of
the PN  has largely merged with the ISM and thus is the key to solve the puzzle of  missing mass.

Despite the points mentioned above favoring the PN-ISM interaction, the possibility
of the \ion{H}{1}  features resulting from contamination of a foreground cloud remains.
The large swept ISM mass is extraordinary for a PN, but is typical for supernova 
remnants (SNRs) and  Wolf-Rayet stars. We notice that indeed a SNR,
G65.2+5.7 ($\alpha _{\rm J2000} = 19^h 33^m$, $\delta _{\rm J2000} = 30\arcdeg 30\arcmin$),
is coincident in position with BD+30. G65.2+5.7 has a diameter of
$\sim 4\arcdeg$ and a distance of 800\,pc \citep{2009A&A...503..827X}, and
thus is likely to contribute to the \ion{H}{1} absorption and emission.  
Additional support for this possibility comes from \citet{2011AAS...21725625V} who
discovered emission from an expanding \ion{H}{1} shell around  G65.2+5.7 within the velocity range
of $-12$\,km\,s$^{-1}$ to 10\,km\,s$^{-1}$. This is consistent with the components C6 and C7, and these could be part of the SNR shell. 
 The \ion{H}{1} emission found towards BD+30 by  \citet{1990ApJ...351..515T}  also falls within this velocity range, but is likely associated with BD+30 as it is spatially unresolved and coincident with BD+30.  The BD+30 emission  measured by \citet{1990ApJ...351..515T}  may  be related to the spatially extended component
detected by FAST, supporting some association between the wide  \ion{H}{1} emission and the PN and the idea that the majority of the \ion{H}{1} emission would be swept ISM material. The presence of large scale emission from the SNR complicates the interpretation for velocities redward of the systemic velocity of BD+30.

\section{Conclusion} \label{sec:concl}

We have searched for atomic hydrogen surrounding the young PN BD+30 using the FAST 19-beam observations. 
The new high-quality spectra confirm the existence 
of \ion{H}{1} emission  toward this PN, as previously discovered by interferometry
\citep{1990ApJ...351..515T}. 
However, we find that the \ion{H}{1} emission
detected by \cite{1990ApJ...351..515T} is in fact resolved into two 
individual velocity components and extends to an angular radius of $\sim 
12\arcmin$ rather than confined to a compact region.
The mass of the \ion{H}{1} shell is larger than 100\,M$_\sun$, excluding
the possibility that it is primarily composed of ejected stellar materials from the host star of BD+30.
A reasonable conjecture is that the large  \ion{H}{1} shell stems from
material swept up by the BD+30-ISM interaction processes at earlier stages of evolution
although such a large swept ISM mass has never been reported for a PN. The \ion{H}{1} emission is enhanced
roughly along the leading edge of BD+30, providing further evidence supporting this  conjecture.

We tentatively detect circumstellar \ion{H}{1}
absorption features. The \ion{H}{1} absorption is expected at various velocities: systemic if it is located in a stalled asterosphere where it sweeps up interstellar gas or at the blueshifted velocities if just outside the ionized region. 
 The \ion{H}{1} gas producing
the absorption is estimated to have a  mass of   $\sim3\times10^{-2}$\,M$_\sun$.

However, we cannot firmly rule out
the possibility that the detected \ion{H}{1} absorption and emission may originate from
a SNR that overlaps along the line of sight and happens to have
similar velocities to BD+30.
If this was the case, the number of PNe with detected \ion{H}{1} emission would
be reduced from five to four.  Nevertheless,
this possibility is not particularly likely 
because it does not seem to fit the interferometric
observations of \cite{1990ApJ...351..515T}.

Investigating PN-ISM interactions in cases like BD+30 is a promising avenue to solve the `PN missing mass problem'. 
The resultant circumstellar structures are prone to be
revealed by the observations of \ion{H}{1} emission.
Unlike interferometric observations that are insensitive to structures larger than a certain extent, single-dish 
telescopes have noteworthy benefits in observing extended 
circumstellar materials. 
Although the results of this study
are somewhat inconclusive, we demonstrate the feasibility of utilizing the
\ion{H}{1} observations of FAST
to efficiently reveal the structures for future PN-ISM interaction studies.

\begin{acknowledgments}

We acknowledge an anonymous reviewer for helpful
suggestions that improve the quality of this paper.
We thank Bi-Wei Jiang and Ke Wang for helpful comments.
The financial supports of this work are from
 the National Science Foundation of China (NSFC, Grant No. 11973099) and the science research grants from the China Manned Space Project (NO. CMS-CSST-2021-A09 and CMS-CSST-2021-A10). 
 QAP thanks the Hong Kong Research Grants Council for GRF research support under grants 17326116 and 17300417.
AAZ acknowledges funding from the UK STFC through grant ST/T000414/1. This work made use of the data from FAST (Five-hundred-meter Aperture Spherical radio Telescope).  FAST is a Chinese national mega-science facility, operated by National Astronomical Observatories, Chinese Academy of Sciences.
This publication utilizes data from Galactic ALFA \ion{H}{1} (GALFA \ion{H}{1}) survey data set obtained with the Arecibo L-band Feed Array (ALFA) on the Arecibo 305m telescope. The Arecibo Observatory is operated by SRI International under a cooperative agreement with the National Science Foundation (AST-1100968), and in alliance with Ana G. M{\'e}ndez-Universidad Metropolitana, and the Universities Space Research Association. The GALFA HI surveys have been funded by the NSF through grants to Columbia University, the University of Wisconsin, and the University of California.

\end{acknowledgments}

%

\vspace{5mm}
\facilities{FAST}


\software{Astropy \citep{2013A&A...558A..33A,2018AJ....156..123A}; LIMFIT \citep{2021zndo...5570790N}}

\bibliography{sample631}{}
\bibliographystyle{aasjournal}




\begin{deluxetable*}{@{\extracolsep{4pt}}crrrrrrrr@{}}[htb]
\tabletypesize{\scriptsize}
\tablewidth{0pt}
\linespread{0.8}
\tablecaption{Properties of C6 and C7 obtained from spectral decomposition. \label{tab:fitting result}}
\tablehead{\\
\colhead{Beam} & \multicolumn{4}{c}{C6} & \multicolumn{4}{c}{C7} \\
\cline{2-5} 
\cline{6-9}
\colhead{} & \colhead{$\int S_v dv$} & \colhead{$V_{\rm p}$} & \colhead{FWHM} & \colhead{$S_{\rm p}$} & \colhead{$\int S_v dv$} & \colhead{$V_{\rm p}$} & \colhead{FWHM} & \colhead{$S_{\rm p}$}\\
\colhead{} & \colhead{(mJy\,km\,s$^{-1}$)} & \colhead{(km\,s$^{-1}$)} & \colhead{(km\,s$^{-1}$)} & \colhead{(mJy)} & \colhead{(mJy\,km\,s$^{-1}$)} & \colhead{(km\,s$^{-1}$)} & \colhead{(km\,s$^{-1}$)} & \colhead{(mJy)} 
}
\startdata
    M01   & 2839$\pm$63 & $-$10.63$\pm$0.02 & 5.34$\pm$0.05 & 400$\pm$7 & 3514$\pm$151 & $-$4.33$\pm$0.01 & 4.16$\pm$0.05 & 635$\pm$22 \\
    M02   & 8059$\pm$786 & $-$6.50$\pm$0.25 & 9.18$\pm$0.19 & 660$\pm$54 & 2152$\pm$213 & $-$4.07$\pm$0.02 & 3.37$\pm$0.08 & 479$\pm$39 \\
    M03   & 8803$\pm$686 & $-$5.87$\pm$0.27 & 11.97$\pm$0.17 & 553$\pm$37 & 13508$\pm$2109 & 1.00$\pm$0.43 & 8.82$\pm$0.30 & 1151$\pm$149 \\
    M04   & 7145$\pm$249 & $-$7.49$\pm$0.08 & 10.63$\pm$0.10 & 505$\pm$14 & 8015$\pm$271 & $-$0.49$\pm$0.03 & 6.22$\pm$0.05 & 969$\pm$29 \\
    M05   & 2143$\pm$41 & $-$11.66$\pm$0.01 & 3.29$\pm$0.03 & 489$\pm$7 & 16076$\pm$620 & $-$4.41$\pm$0.09 & 12.37$\pm$0.11 & 977$\pm$31 \\
    M06   & 7529$\pm$213 & $-$9.06$\pm$0.06 & 7.81$\pm$0.07 & 725$\pm$16 & 3597$\pm$136 & $-$4.55$\pm$0.03 & 4.55$\pm$0.05 & 593$\pm$18 \\
    M07   & 6015$\pm$459 & $-$8.09$\pm$0.27 & 10.73$\pm$0.22 & 421$\pm$26 & 4631$\pm$146 & $-$4.82$\pm$0.01 & 3.57$\pm$0.03 & 974$\pm$25 \\
    M08   & 2599$\pm$277 & $-$9.82$\pm$0.14 & 6.20$\pm$0.20 & 315$\pm$26 & 6119$\pm$377 & $-$4.63$\pm$0.04 & 5.15$\pm$0.08 & 893$\pm$46 \\
    M09   & 3678$\pm$473 & $-$8.65$\pm$0.27 & 7.91$\pm$0.27 & 350$\pm$36 & 4140$\pm$255 & $-$4.52$\pm$0.05 & 4.61$\pm$0.08 & 675$\pm$34 \\
    M10   & 7418$\pm$394 & $-$6.52$\pm$0.13 & 8.93$\pm$0.13 & 625$\pm$27 & 1456$\pm$128 & $-$3.04$\pm$0.03 & 3.05$\pm$0.08 & 359$\pm$25 \\
    M11   & 4678$\pm$289 & $-$9.20$\pm$0.08 & 6.90$\pm$0.10 & 510$\pm$26 & 12073$\pm$2808 & $-$1.00$\pm$0.48 & 9.57$\pm$0.51 & 948$\pm$182 \\
    M12   & 11705$\pm$379 & $-$6.80$\pm$0.13 & 11.03$\pm$0.10 & 798$\pm$21 & 2896$\pm$1013 & $-$1.60$\pm$0.33 & 5.64$\pm$0.36 & 386$\pm$116 \\
    M13   & 3756$\pm$179 & $-$10.85$\pm$0.03 & 4.98$\pm$0.08 & 567$\pm$21 & 16958$\pm$1332 & $-$1.22$\pm$0.27 & 13.61$\pm$0.42 & 937$\pm$52 \\
    M14   & 4706$\pm$63 & $-$11.59$\pm$0.01 & 3.69$\pm$0.02 & 957$\pm$9 & 16024$\pm$520 & $-$3.48$\pm$0.09 & 11.81$\pm$0.15 & 1020$\pm$24 \\
    M15   & 2000$\pm$81 & $-$10.07$\pm$0.02 & 3.69$\pm$0.05 & 407$\pm$13 & 12822$\pm$186 & $-$6.80$\pm$0.04 & 9.41$\pm$0.04 & 1025$\pm$12 \\
    M16   & 3616$\pm$349 & $-$11.16$\pm$0.20 & 7.56$\pm$0.19 & 359$\pm$28 & 4935$\pm$623 & $-$5.12$\pm$0.11 & 7.51$\pm$0.29 & 494$\pm$48 \\
    M17   & 684$\pm$77 & $-$13.07$\pm$0.13 & 4.91$\pm$0.26 & 105$\pm$8 & 7502$\pm$234 & $-$5.46$\pm$0.03 & 7.03$\pm$0.07 & 802$\pm$21 \\
    M18   & 253$\pm$68 & $-$11.27$\pm$0.19 & 3.09$\pm$0.41 & 62$\pm$12 & 5870$\pm$100 & $-$5.51$\pm$0.01 & 3.82$\pm$0.03 & 1155$\pm$15 \\
    M19   & 3195$\pm$151 & $-$9.50$\pm$0.14 & 8.88$\pm$0.17 & 270$\pm$9 & 8497$\pm$142 & $-$4.34$\pm$0.01 & 4.64$\pm$0.02 & 1375$\pm$18 \\
\enddata
\end{deluxetable*}

\end{document}